**Plasma Fixed Nitrogen (PFN) Improves Lettuce Field Holding Potential**

Benjamin Wang[1], Qiyang Hu[1], Bruno Felix Castillo[1], Christina Simley[1], Andrew Yates[1], Brian Sharbono[2], Kyle Brasier[3], Mark A. Cappelli[1]

[1] Department of Mechanical Engineering, Stanford University, Stanford CA 94305

[2] Stanford Woods Institute for the Environment, Stanford University, Stanford CA 94305

[3] Vilmorin-Mikado USA Inc., Salinas, CA 93901

ADDITIONAL INDEX WORDS. *plasma fixed nitrogen, romaine lettuce, nitrogen efficiency, agronomic quality, marketable yield, bolting prevention, conventional treatment, control group, stem quality, leaf turgidity*

This project was funded by the Stanford Sustainability Accelerator, Stanford High Impact Technology (HIT) Fund, Stanford Woods Institute Realizing Environmental Innovation Program, and the Stanford TomKat Innovation Transfer Grant.

**Summary**

We show that plasma fixed nitrogen (PFN) as a biostimulant can improve marketable lettuce yield following delayed harvest. Using just one-tenth of the conventional nitrogen, PFN, generated by a dielectric barrier discharge over water, was field-tested against traditional fertilization methods. PFN increased marketable lettuce yield by 250% over conventional growing methods despite reducing applied nitrogen fertilizer. Our study suggests that PFN can increase marketable lettuce yield following delayed harvest while ensuring environmental sustainability and product quality.

**Introduction**

One of the challenges in conventional agriculture is the reliance on nitrogen fertilizers. Its overuse leads to environmental degradation, eutrophication of water bodies, and in-field nitrous oxide emissions. The energy-intensive fertilizer synthesis and their transportation contributes to resource depletion and $CO_2$ emissions. The emergence of plasma technology enabling the synthesis of plasma fixed nitrogen (PFN) from renewable electricity presents a revolutionary shift towards sustainable farming. Low temperature plasmas have the capability to activate water and fixate atmospheric nitrogen, creating a reactive solution rich in nitrates and other compounds beneficial for plant growth (Bradu et al. 2020). PFN harnesses the efficient ionization and bond-breaking capability of plasmas, infusing water with nitrogenous compounds, hydroxyl radicals, and hydrogen peroxide, all known to enhance seed germination, root development, and overall plant vitality (Thirumdas et al. 2018). Simultaneously, the PFNs offer a novel approach to utilizing the abundant nitrogen in the atmosphere, transforming it into a form readily used by plants, that can lessen the use of conventional fertilizers.

This short note describes a field study of the use of PFN as a biostimulant for romaine lettuce grown in the Salinas Valley (SV). SV is a major region of lettuce heart production. This region's groundwater has high nitrate levels (over 20 ppm NO3-N), largely attributed to chemical-intensive agricultural practices. Lettuce growers typically apply between 130 and 200 lbs. of N per acre which often surpasses crop needs, to hedge against under fertilization and reduced marketable yields (Cahn et a. 2016). Several studies have explored the use of plasma-generated nitrates as a substitute for traditional nitrogen sources, focusing on direct nutrient supplementation (Subramanian et al. 2021, Ruamrungsri et al. 2023). Previous research has demonstrated that PFN serves as an efficacious source of nitrates for turfgrass growth, with promising outcomes in growth rate and overall plant health (Sze et al. 2021).

**Materials and Methods**

Air plasmas were generated using an atmospheric dielectric barrier discharge (DBD) reactor operating at 23 kHz driving frequency and with a sinusoidal peak-to-peak voltage of 7 kV. The discharge draws 1.8 kW of input power. Industrial water in a recirculating channel generates a free-surface flow close to the surface of the DBD and is treated until the PFN solution reaches a desired pH and nitrate level. The nitrate concentration (as nitrates - $NO_3$–N) of the PFN solutions was measured in-situ using an ion-selective electrode (Vernier GDX-NO3). The pH was measured using a glass bodied pH sensor (Vernier GDX-GPH). These values, as well as the overall composition of the solutions were confirmed by sample analyses by an outside analytical laboratory with A2LA and ISO certification.

Romaine lettuce beds were established as a randomized complete block design on standard two row plots measuring 1.0 m wide and 7.6 m long. Transplants were planted on August 17, 2023, in Salinas, CA (36°37′ N, 121°33′ W) and harvested on October 14, 2023. The study relied on drip irrigation, kept constant between treatments, to meet water demands and followed a mixed species cover crop containing faba bean (*Vicia faba*), pea (*Pisum sativum*), hairy vetch (*Vicia villosa*), and oat (*Avena sativa*). The hearting variety, 'Vicious', was selected for the trial due to its popularity in organic agriculture.

Lettuce plots were subjected to three treatments: a zero-fertilizer control, conventional fertilizer (calcium ammonium nitrate [17-0-0]) (CN), and plasma fixed nitrogen (PFN) with four replications per treatment. No added fertilizer was applied to the zero-fertilizer control treatment while the conventional treatment received 75 kg N/ha and PFN treatment received 8 kg N/ha split over three applications between September 21st and October 5th. Irrigation water contained 40 mg/l of nitrogen while soils contained 41 mg/l of nitrogen at the time of transplanting.

Phenotypic data was collected seven days after the optimal harvest date for romaine hearts to observe field holding capacity. Six randomly selected lettuce plants were selected from each plot and chopped for hearts - defined as the inner portion of the lettuce plant where leaves are folded to form an enclosed structure. Romaine heart fresh weight, core quality, and height were measured. R (4.3.1) was used to perform Analysis of Variance on the data for statistical significance of measurements.

**Results and Discussion**

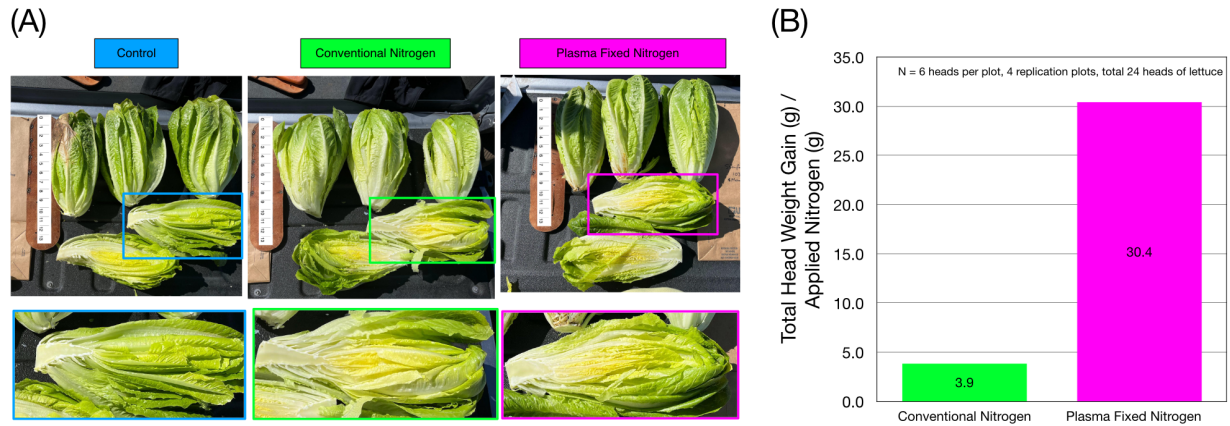

**Figure 1.** (A) Representative images of romaine hearts grown under no fertilizer control, conventional fertilizer, and plasma fixed nitrogen treatments. (B) Plot of the total head weight gain per unit of applied nitrogen for conventional nitrogen and plasma fixed nitrogen treatments.

Figure 1(A) shows harvested lettuce samples with longitudinal cross sections. The control group exhibited signs indicative of suboptimal quality derived from core elongation (i.e. early bolting with increased spacing between leaves), which reduced marketable quality of the lettuce. In contrast, these issues were less pronounced in the lettuce from the conventional nitrogen treatment. While the PFN treatment maintained some indications of bolting, leaf gaps, and reduced turgidity, they were noticeably reduced. However, lettuce grown under the PFN treatment exhibited greater heart quality, with minimal signs of bolting or leaf gaps. The leaves were particularly turgid and high in density, indicative of optimal hydration and cell structure. In an evaluation of nitrogen use efficiency (NUE) across different treatment trials, the PFN treatment produced greater marketable yield per unit of applied nitrogen compared to the conventional conditions - likely due to the high concentrations of nitrogen in water and soil. Figure 1(B) plots the ratio of mass gain over applied nitrogen. For every gram of applied nitrogen, the CN treatment group resulted in a head weight gain of approximately 3.9 grams. The PFN treatment achieved a head weight gain of 30.4 grams per gram of applied nitrogen.

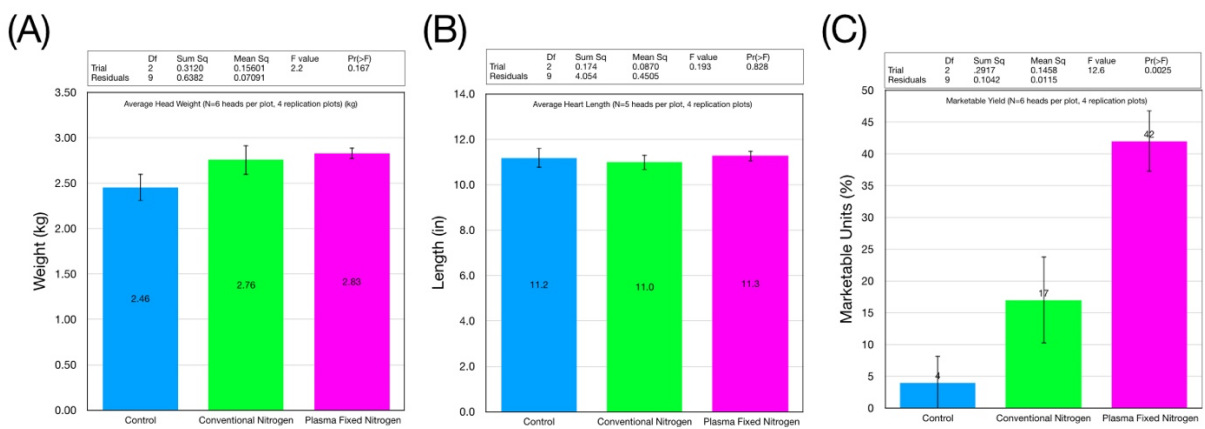

**Figure 2.** Plots showing (A) the average heart weight, (B) heart length, and (C) percentage of marketable romaine hearts for the "Control (No Fertilizer)", "Conventional Nitrogen" (CN), and "Plasma Fixed Nitrogen" (PFN) treatment groups. All error bars represent the standard error of the mean SEM ($\sigma/\sqrt{n}$).

Figure 2(A) illustrates the average heart weight where the PFN treatment averaged 2.83 kg, which was not significantly greater than the CN treatment (2.75 kg). Despite its low percentage of marketable romaine hearts, the control treatment had an average weight of 2.45 kg. The Standard Error of the Mean (SEM) of the PFN group (0.06 kg) is lower than the control (0.15 kg) and conventional nitrogen group (0.17 kg), suggesting more consistency. This finding is consistent with previous experiments in the Salinas Valley where lettuce yields do not significantly differ when more than 100 kg N/ha is available from water, soil, and fertilizer (Hoque et al., 2010). Figure 2(B) plots the average length of the hearts across the treatment groups, with no statistical difference. Figure 2(C) plots the marketable yield across the three treatment groups following the forementioned methodology. The PFN treatment demonstrated the highest marketable heart yield seven days after optimal harvest date (42%) followed by the conventional (17%) and control (4%) treatments.

The interaction of plasma-activated species with plant roots can stimulate root growth and density, increasing the surface area available for nutrient uptake (Ka et al. 2021). As a result, plants treated with PFN not only have access to a more readily available form of nitrogen, but also exhibit enhanced capability to draw this essential nutrient from their environment.

We believe that this is the first field study demonstrating PFN as a biostimulant that increased the field holding capacity of lettuce. The complex interplay between PFN and plant physiology, especially in the realm of root growth and nutrient uptake, necessitates further research to integrate these insights into a comprehensive model for maximizing crop yield and quality.